\documentclass[%
 reprint,
 amsmath,amssymb,
 aps,
 pre,
]{revtex4-1}

\usepackage{pgfplots}
\usetikzlibrary{arrows}
\usetikzlibrary{automata,positioning}
\usepgfplotslibrary{fillbetween}
\usepackage[american]{circuitikz} 
\makeatletter
\def\intkern@{\mkern-8mu\mathchoice{\mkern-4mu}{}{}{}}
\makeatother
\usepackage{graphicx}% Include figure files
\usepackage{dcolumn}% Align table columns on decimal point
\usepackage{bm}% bold math
\begin{document}

\preprint{APS/123-QED}

\title{Statistical mechanical constitutive theory of polymer networks:\\
the inextricable links between distribution, behavior, and ensemble}
%the effects of consistency in modeling}
% Force line breaks with \\
%\thanks{A footnote to the article title}%

\author{Michael R. Buche}
 %\altaffiliation[Also at ]{Physics Department, XYZ University.}%Lines break automatically or can be forced with \\
\author{Meredith N. Silberstein}%
 \email{ms2682@cornell.edu}
\affiliation{%
Theoretical and Applied Mechanics, \\ 
 Sibley School of Mechanical and Aerospace Engineering, \\ Cornell University, Ithaca, New York 14853
}%

%\collaboration{MUSO Collaboration}%\noaffiliation

%\author{Charlie Author}
 %\homepage{http://www.Second.institution.edu/~Charlie.Author}
%\affiliation{
 %Second institution and/or address\\
 %This line break forced% with \\
%}%
%\affiliation{
 %Third institution, the second for Charlie Author
%}
%\author{Delta Author}
%\affiliation{%
 %Authors' institution and/or address\\
 %This line break forced with \textbackslash\textbackslash
%}%

% \collaboration{CLEO Collaboration}%\noaffiliation

\date{\today}% It is always \today, today,
             %  but any date may be explicitly specified

\begin{abstract}
%An article usually includes an abstract, a concise summary of the work covered at length in the main body of the article. 

A fundamental theory is presented for the mechanical response of polymer networks undergoing large deformation which seamlessly integrates statistical mechanical principles with macroscopic thermodynamic constitutive theory.
Our formulation permits the consideration of arbitrary polymer chain behaviors when interactions among chains may be neglected.
This careful treatment highlights the naturally occurring correspondence between single-chain mechanical behavior and the equilibrium distribution of chains in the network, as well as the correspondences between different single-chain thermodynamic ensembles. 
We demonstrate these important distinctions with the extensible freely jointed chain model.
This statistical mechanical theory is then extended to the continuum scale, where we utilize traditional macroscopic constitutive theory to ultimately retrieve the Cauchy stress in terms of the deformation and polymer network statistics.
Once again using the extensible freely jointed chain model, we illustrate the importance of the naturally occurring statistical correspondences through their effects on the stress-stretch response of the network.
We additionally show that these differences vanish when the number of links in the chain becomes sufficiently large enough, and discuss why certain methods perform better than others before this limit is reached.  
\\

\noindent DOI: \href{https://doi.org/10.1103/PhysRevE.102.012501}{10.1103/PhysRevE.102.012501}.

%
%\begin{description}
%\item[Usage]
%Secondary publications and information retrieval purposes.
%\item[Structure]
%You may use the \texttt{description} environment to structure your abstract; use the optional argument of the \verb+\item+ command to give the category of each item. 
%\end{description}
\end{abstract}

%\keywords{Suggested keywords}%Use showkeys class option if keyword
                              %display desired
\maketitle

%\tableofcontents

%%%%%%%%%%%%%%%%%%%%%%%%%%%%%%%%%%%%%%%%%%%%%%%%%%%%%%%%%%
%%%%%%%%%%%%%%%%%%%%%%%%%%%%%%%%%%%%%%%%%%%%%%%%%%%%%%%%%%
\section{\label{sec1}Introduction}

Understanding the mechanics of polymer networks is important for improving and predicting the mechanical behaviors of a wide range of polymeric materials, from physically-crosslinked rubbers to mechanochemically-responsive networks.
Constitutive models that are grounded in statistical mechanics are especially useful because they allow the direct incorporation of molecular phenomena and thus a fundamental understanding of the material.
In order to establish a model with such predictive power, one needs to proceed from the statistical mechanics of single polymer chains all the way to the macroscopic mechanical behavior of the entire material.
The meticulous detail we offer here is required to retain generality throughout this process.

Polymer network constitutive models often utilize a single polymer chain statistical mechanical model. 
The most common of these single chain models is the freely jointed chain (FJC) model \cite{doi1988theory}, where rigid links are connected in series and allowed to rotate about the connecting hinges without change in energy. 
The mechanical response of the FJC under end-to-end extension is determined by the reduction in configurational entropy, which is then directly connected to the equilibrium probability distribution of end-to-end lengths by Boltzmann's entropy formula \cite{kuhn1942beziehungen}. 
For a large number of links or in the case of an applied force, the mechanical response and probability distribution of end-to-end lengths may be written analytically using the inverse Langevin function \cite{treloar1949physics}.
In the case of an applied extension, more sophisticated methods are necessary to obtain the mechanical response and distribution of end-to-end lengths, such as those using series expansions \cite{wang1952statistical} or those that transform between thermodynamic ensembles \cite{manca2012elasticity}.
When the number of links approaches infinity, the probability distribution of end-to-end lengths obeys Gaussian statistics \cite{flory1974moments,yoon1974moments}, and for end-to-end length much smaller than the contour length, the mechanics of the chain become that of the ideal, linear chain \cite{flory1979molecular}.
When the end-to-end length approaches the contour length, the FJC becomes infinitely stiff due to its inextensibility.
The FJC model can be expanded to that of the extensible freely jointed chain (EFJC) by replacing the rigid links with stiff harmonic springs \cite{fiasconaro2019analytical}.
Now when the end-to-end length approaches the contour length, the EFJC begins to stretch the stiff harmonic links and subsequently achieves end-to-end lengths greater than the original contour length, hence it is extensible.
Another popular set of models are the freely rotating chain (FRC) models, where the FJC model is adjusted by fixing all bond angles and only permitting torsional angles to freely rotate \cite{rubinstein2003polymer}.
This model cannot be solved analytically and therefore requires careful numerical techniques \cite{livadaru2003stretching}.
For small bond angles, the FRC model becomes the Kratky-Porod (or discrete worm-like chain) model \cite{kratky1949rntgenuntersuchung}, and when the link length additionally becomes small compared to the contour length, the FRC model becomes the continuous worm-like chain (WLC) model. 
Both the discrete and continuous forms of the WLC model have been expanded to include stiff harmonic springs \cite{kierfeld2004stretching}.
Recent single chain constitutive models include covalent bond rupture \cite{mao2017rupture} and mechanochemically-activated bonds \cite{lavoie2019modeling}.
It is apparent from this vast literature that the correspondences between end-to-end length probability distribution, the mechanical behavior, and the applied boundary conditions (thermodynamic ensemble) are of vital significance. 

Upon establishing a statistical description by way of single chain mechanics, the model derivation must then proceed to connect the macroscopic deformation of the polymer network to this single chain description. 
This is typically accomplished by way of constructing the Helmholtz free energy density, prescribing some aspect of the network evolution in terms of the macroscopic deformation, connecting the network evolution back to individual chains, and using 2nd law of thermodynamics analysis. 
Most often the 2nd law analysis results in a hyperelastic model, which means that the stress is directly related to the derivative of the free energy density with respect to the deformation gradient. 
After choosing a single chain model, the construction of the free energy density for the network involves the choice of the distribution of initial chain lengths and orientations in the network. 
Several models have used discretely-oriented chains to represent the distribution of chains in the network, such as the 3-chain \cite{james1943theory,wang1952statistical}, 4-chain \cite{flory1943statistical,treloar1946elasticity}, 8-chain \cite{arruda1993three,mao2018fracture}, and 21 chain \cite{miehe2004micro,silberstein2014modeling} models. 
Other models utilize a continuous orientation distribution of chains to represent the network, where some assume that all end-to-end lengths are initially the same \cite{treloar1954photoelastic,wu1993improved} and others consider an initial distribution of end-to-end lengths \cite{tanaka1992viscoelastic,storm2005nonlinear,vernerey2017statistically}.
Polydispersity, i.e. varying contour lengths, may be included in either the discrete \cite{wang2015mechanics,lavoie2019continuum} or continuous \cite{tehrani2017effect,tehrani2018polydispersity} distribution formulations. 
An affine or non-affine deformation of the distribution can be prescribed, where the non-affinity can be a fundamental aspect of the initial distribution \cite{arruda1993three} or based upon some physical constraint \cite{miehe2004micro,tkachuk2012maximal,verron2017equal}.
Unfortunately, the natural correspondence between the choice of single chain model and the chain length distribution within the network tends to be ignored in these models.

Despite decades of work, there remains a need for a methodical and general statistical mechanics derivation of polymer network mechanics such that the assumptions and their implications are apparent.
The approach taken here begins from fundamental statistical mechanics, makes clear all assumptions and places emphasis on the correspondences between the network distribution, single chain mechanics, and the different thermodynamic ensembles. 
Most preceding constitutive models make no reference to such nuances, and as a result risk making considerable mistakes.
Furthermore, there has been no study on the effects that these correspondences have on the macroscopic mechanical response of the network and when they may be ignored.
Such an approach will also take great care in stitching this general statistical description into the macroscopic description, performing a detailed 2nd law analysis to retrieve the stress and making sure that all neglected terms are truly negligible. 
Many preceding constitutive models use considerable assumptions to construct the stress, lose generality by choosing a specific chain model during the 2nd law analysis, and/or neglect terms that could contribute to the stress without proof that they can be neglected.

In this manuscript, we present a constitutive model for polymer networks undergoing finite deformation that is constructed with great detail. 
We begin in Section~\ref{sec2.1} from fundamental statistical mechanics, ensuring that the correspondences between the distribution of chains in the network and the mechanics of single chains are understood and accounted for. 
We also account for the differences between thermodynamic ensembles, ensuring we utilize the correct ensemble and understand the correspondence relations that allow us to go from one to the other.
Prescribing an affine deformation to the network distribution, we extend the statistical theory to the macroscale and perform a detailed 2nd law analysis in order to retrieve the stress in Section~\ref{sec2.2}. 
In the process, we perform many mathematical manipulations in order to maintain the generality of the model.
This includes the proof that a term produced when integrating by parts is indeed zero for relevant chain models, which has been previously taken for granted.
With this detailed framework, in Section~\ref{sec3} we are able to study the macroscopic effects of the aforementioned statistical correspondences and show that they are considerable when chains are not sufficiently long.
Throughout the manuscript, the EFJC model is used to demonstrate the statistical correspondences within the network and their effects on the macroscopic mechanical response of the network. 
This framework will prove useful in constructing future constitutive models for more complicated polymer networks.

%%%%%%%%%%%%%%%%%%%%%%%%%%%%%%%%%%%%%%%%%%%%%%%%%%%%%%%%%%
\begin{figure*}[t]
\begin{center}
\includegraphics{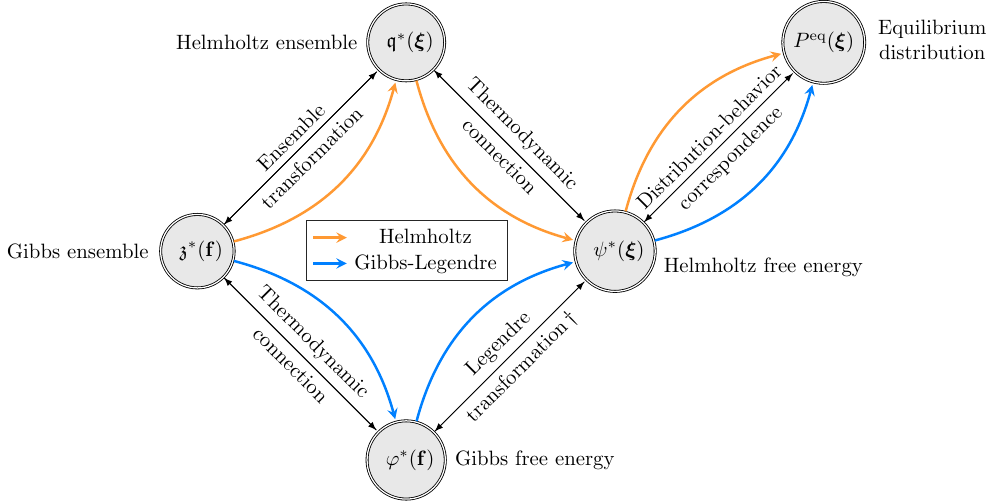}
\end{center}
\caption{\label{scheme1}
  Diagram describing the exact (Helmholtz) and approximate (Gibbs-Legendre) methods of arriving at the single chain Helmholtz free energy $\psi^*(\boldsymbol{\xi})$ of a chain with end-to-end vector $\boldsymbol{\xi}$ and equilibrium probability distribution $P^\mathrm{eq}(\boldsymbol{\xi})$ of chains with that end-to-end vector. 
  $\dagger$\,The Gibbs-Legendre method is approximate since the necessary Legendre transformation is only valid in the thermodynamic limit of long chains.
}
\end{figure*}
%%%%%%%%%%%%%%%%%%%%%%%%%%%%%%%%%%%%%%%%%%%%%%%%%%%%%%%%%%

%%%%%%%%%%%%%%%%%%%%%%%%%%%%%%%%%%%%%%%%%%%%%%%%%%%%%%%%%%
%%%%%%%%%%%%%%%%%%%%%%%%%%%%%%%%%%%%%%%%%%%%%%%%%%%%%%%%%%
\section{\label{sec2}General theory}

%%%%%%%%%%%%%%%%%%%%%%%%%%%%%%%%%%%%%%%%%%%%%%%%%%%%%%%%%%
\subsection{\label{sec2.1}Statistical mechanical description}

Here we present a statistical mechanical description of an ensemble of noninteracting polymer chains. 
The statistical mechanical description naturally provides explicit relationships between the equilibrium distribution of polymer chain end-to-end vectors and the Helmholtz free energy of a polymer chain with a given end-to-end vector; we refer to these as the distribution-behavior correspondence relations.
The original thermodynamic ensemble (Helmholtz) for a single chain is parameterized by an end-to-end vector, where a simple Laplace transformation allows parameterization by the end-to-end force in another ensemble (Gibbs).
Since we desire results from the Helmholtz ensemble but are often only able to compute the partition function in the Gibbs ensemble, ensemble transformation relations between the two -- both exact and in the thermodynamic limit -- are provided.
When obtaining the single chain free energy function and the equilibrium distribution of end-to-end vectors, we refer to the method that utilizes the exact transformation as the Helmholtz method, and that utilizing the transformation in the thermodynamic limit as the Gibbs-Legendre method; see Fig.~\ref{scheme1} for a schematic.  
Next, we illustrate these features of our statistical description using the EFJC model as an example chain model. 
We complete the statistical theory by formulating the general evolution law for the polymer network distribution of end-to-end vectors.

\subsubsection{Helmholtz ensemble}

The polymer network is taken to be represented by an ensemble of $N$ indistinguishable noninteracting polymer chains. The canonical partition function is then

\begin{equation}\label{Qeqn}
	Q = 
	\frac{\mathfrak{q}^N}{N!}
	,
\end{equation}
where the single chain partition function $\mathfrak{q}$ is given by a classical integration over the coordinates $\mathbf{q}_j$ and momenta $\mathbf{p}_j$ of each of the $M$ atoms in the chain \cite{mcq},

\begin{equation}\label{qeqn}
	\mathfrak{q} = 
	\frac{1}{h^{3M}}\idotsint e^{-\beta\varepsilon} \prod_{j=1}^{M} d^3\mathbf{p}_j\,d^3\mathbf{q}_j
	.
\end{equation}
Here $h$ is Planck's constant and $\beta=1/kT$ is the inverse temperature, with Boltzmann's constant $k$ and temperature $T$. 
The Hamiltonian $\varepsilon$ of the chain is

\begin{equation}
	\varepsilon = 
	{u}(\mathbf{q}_1,\ldots,\mathbf{q}_{M}) + \sum_{j=1}^{M} \frac{p_j^2}{2m_j}
	,
\end{equation}
where ${u}$ is the potential energy function describing interaction energies between atoms within the polymer chain, and $m_j$ is the mass of $j^\mathrm{th}$ atom in the chain.
The momentum integrations are completed to write a portion of the chain partition function:

\begin{equation}
	\mathfrak{q}_\mathrm{mom} 
	= 
	\prod_{j=1}^{M} \left(\frac{2\pi m_j kT}{h^2}\right)^{3/2}
	\label{qmomentum}
	.
\end{equation}
If we take the atomic coordinates relative to the first atom along the chain backbone, $\mathbf{r}_j=\mathbf{q}_j-\mathbf{q}_1$, we can complete the rigid body translation integration -- where the chain is translated over the whole volume -- and pick up a factor of $V$.
We now have

\begin{equation}
	\mathfrak{q} = 
	\mathfrak{q}_\mathrm{con} \mathfrak{q}_\mathrm{mom} V
	,
\label{qqqeqn}
\end{equation}
where the chain configuration integral $\mathfrak{q}_\mathrm{con}$ is then

\begin{equation}
	\mathfrak{q}_\mathrm{con} = 
	\idotsint e^{-\beta{u}} \prod_{j=2}^{M} d^3\mathbf{r}_j
	.
\end{equation}
If the $M^\mathrm{th}$ atom is the last atom along the chain backbone, we seek to calculate the probability density distribution $P^\mathrm{eq}(\boldsymbol{\xi})$ that a chain has the end-to-end vector $\mathbf{r}_{M}=\boldsymbol{\xi}$ at equilibrium.
This means that the probability that a chain has the end-to-end vector within $d^3\boldsymbol{\xi}$ of $\boldsymbol{\xi}$ at equilibrium would be $P^\mathrm{eq}(\boldsymbol{\xi})\,d^3\boldsymbol{\xi}$. 
We then write $\mathfrak{q}^*$, the chain configuration integral corresponding to end-to-end vector $\boldsymbol{\xi}$, by integrating the Dirac delta function $\left(\delta\right)$

\begin{eqnarray}
	\mathfrak{q}^*(\boldsymbol{\xi}) &&= 
	\idotsint e^{-\beta{u}(\mathbf{r}_2,\ldots,\mathbf{r}_M)} \delta^3\left(\mathbf{r}_M - \boldsymbol{\xi}\right) \prod_{j=2}^{M} d^3\mathbf{r}_j
	,\\
	&&=
	\idotsint e^{-\beta{u}(\mathbf{r}_2,\ldots,\boldsymbol{\xi})} \prod_{j=2}^{M-1} d^3\mathbf{r}_j
	.
\label{qfullconfig}
\end{eqnarray}
According to Boltzmann statistics, the probability of a single chain configuration at thermodynamic equilibrium is $e^{-\beta u}/\mathfrak{q}_\mathrm{con}$, so we integrate over all configurations that have the end-to-end vector $\boldsymbol{\xi}$ in order to retrieve $P^\mathrm{eq}(\boldsymbol{\xi})$,

\begin{eqnarray}
	P^\mathrm{eq}(\boldsymbol{\xi}) &&= 
	\idotsint \frac{e^{-\beta{u}}}{\mathfrak{q}_\mathrm{con}}\,\delta^3\left(\mathbf{r}_M - \boldsymbol{\xi}\right) \prod_{j=2}^{M} d^3\mathbf{r}_j
\label{MCMCcan}
	,\\
	&&=
	\frac{\mathfrak{q}^*(\boldsymbol{\xi})}{\iiint \mathfrak{q}^*(\tilde{\boldsymbol{\xi}})\,d^3\tilde{\boldsymbol{\xi}}}
	=
	\frac{\mathfrak{q}^*(\boldsymbol{\xi})}{\mathfrak{q}_\mathrm{con}}
	,
\label{equiv0}
\end{eqnarray}
where $\tilde{\boldsymbol{\xi}}$ is a dummy variable of integration; the tilde will continue to denote dummy variables of integration. 
If this equilibrium distribution is rotationally symmetric (only varies with $\xi=\sqrt{\boldsymbol{\xi}\cdot\boldsymbol{\xi}}$), we can use the equilibrium radial distribution function

\begin{equation}\label{geq}
	g^\mathrm{eq}(\xi) = 4\pi\xi^2 P^\mathrm{eq}(\boldsymbol{\xi})
	.
\end{equation}
The chain Helmholtz free energy $\psi^*$ associated with $\mathfrak{q}^*$ is, from the principal thermodynamic connection formula \cite{mcq},

\begin{equation}
	\psi^*(\boldsymbol{\xi}) = 
	- kT \ln \mathfrak{q}^*(\boldsymbol{\xi})
\label{psifromq}
	,
\end{equation}
so we may finally write the equilibrium distribution as

\begin{equation}
	P^\mathrm{eq}(\boldsymbol{\xi}) = 
	\dfrac{ e^{-\beta\psi^*(\boldsymbol{\xi})} }{ \iiint  e^{-\beta\psi^*(\tilde{\boldsymbol{\xi})}}\,d^3\tilde{\boldsymbol{\xi}} }
	,
\label{equiv1}
\end{equation}
which, if $\psi^*(\boldsymbol{\xi}_\mathrm{ref})=\psi^*_\mathrm{ref}$ is known for some $\boldsymbol{\xi}_\mathrm{ref}$, we have

\begin{equation}
	\psi^*(\boldsymbol{\xi}) = 
	\psi^*_\mathrm{ref} - kT\ln\left[\frac{P^\mathrm{eq}(\boldsymbol{\xi})}{P^\mathrm{eq}(\boldsymbol{\xi}_\mathrm{ref})}\right]
	.
\label{equiv2}
\end{equation}
We refer to Eqs.~\eqref{equiv1} and \eqref{equiv2} as the distribution-behavior correspondence relations, as they show a one-to-one correspondence between the free energy of a chain for a given end-to-end vector and the equilibrium distribution of such end-to-end vectors. 

\subsubsection{Gibbs ensemble}

The Gibbs ensemble releases the end-to-end vector constraint of the Helmholtz ensemble and instead applies an end-to-end force. 
The Helmholtz ensemble coincides with the canonical ensemble (which has the Helmholtz free energy as the principal thermodynamic potential), but the Gibbs ensemble does not exactly coincide with the isobaric-isothermal ensemble (which has the Gibbs free energy as the principal thermodynamic potential), despite an applied force seeming to be analogous to an applied pressure. 
The naming of the Gibbs ensemble is then perhaps a bit misleading, but we will continue to use it since it seems to have become standard.
The Gibbs ensemble Hamiltonian is

\begin{equation}
	\varepsilon = 
	{u}(\mathbf{q}_1,\ldots,\mathbf{q}_{M}) + \sum_{j=1}^{M} \frac{p_j^2}{2m_j}
	-\mathbf{f}\cdot(\mathbf{q}_M-\mathbf{q}_1)
	,
\end{equation}
where $\mathbf{f}$ is the force that acts equally and oppositely on atoms 1 and $M$ at the ends of the polymer chain, and $\mathbf{q}_M-\mathbf{q}_1=\mathbf{r}_M$ is the end-to-end vector of the chain. 
The system partition function and the momentum partition function take the same form as Eqs.~\eqref{Qeqn} and \eqref{qmomentum}, respectively, and we receive the same factor of $V$ from the rigid body translation integration, but the chain configuration integral corresponding to force $\mathbf{f}$,

\begin{eqnarray}
	\mathfrak{z}^*(\mathbf{f}) = &&
	\idotsint e^{-\beta{u}} e^{\beta\mathbf{f}\cdot\mathbf{r}_M} \prod_{j=2}^{M} d^3\mathbf{r}_j
	,\\ = &&
	\iiint \mathfrak{q}^*(\boldsymbol{\xi}) e^{\beta\mathbf{f}\cdot\boldsymbol{\xi}} \,d^3\boldsymbol{\xi}
\label{Laplacetrans1}
	,
\end{eqnarray}
is now utilized.
The chain configuration integral corresponding to the Gibbs ensemble is directly a Laplace transform of the Helmholtz ensemble chain configuration integral. 
The probability density distribution that a chain experiences the force $\mathbf{f}$ at equilibrium $P_\mathfrak{z}^\mathrm{eq}(\mathbf{f})$ is then given by the ratio of $\mathfrak{z}^*(\mathbf{f})$ to the integral of $\mathfrak{z}^*(\mathbf{f})$ over all end-to-end force vectors $\mathbf{f}$.
The principal thermodynamic connection formula yields the Gibbs free energy $\varphi^*$ associated with the force,

\begin{equation}
	\varphi^*(\mathbf{f}) = 
	-kT\ln\mathfrak{z}^*(\mathbf{f})
\label{freeenergygibbs}
	,
\end{equation}
from which we obtain the Gibbs ensemble distribution-behavior correspondence relations:

\begin{eqnarray}
	P_\mathfrak{z}^\mathrm{eq}(\mathbf{f}) &&= 
	\dfrac{ e^{-\beta\varphi^*(\mathbf{f})} }{ \iiint  e^{-\beta\varphi^*(\tilde{\mathbf{f}})}\,d^3\tilde{\mathbf{f}} }
	,
\label{equiv1gibbs}
\\
	\varphi^*(\mathbf{f}) &&= 
	\varphi^*_\mathrm{ref} - kT\ln\left[\frac{P_\mathfrak{z}^\mathrm{eq}(\mathbf{f})}{P_\mathfrak{z}^\mathrm{eq}(\mathbf{f}_\mathrm{ref})}\right]
	.
\label{equiv2gibbs}
\end{eqnarray}

\subsubsection{Ensemble transformations}

It has been demonstrated that the mechanical response of a given polymer chain model can differ appreciably between the two ensembles if the thermodynamic limit (i.e. chains consisting of sufficiently many links) is not satisfied \cite{manca2012elasticity}. 
This is an issue because while traditional macroscopic constitutive theories require the Helmholtz free energy of the system, single polymer chain partition functions are often only solvable, if at all, in the Gibbs ensemble.
It is for this reason we require general formulae to transform one ensemble into the other: Eq.~\eqref{Laplacetrans1} allows one to retrieve the Gibbs ensemble from the Helmholtz ensemble; its inversion, from Manca, et.~al. \cite{manca2012elasticity}, allows one to retrieve the Helmholtz ensemble from the Gibbs ensemble:

\begin{equation}
	\mathfrak{q}^*(\boldsymbol{\xi}) = 
	\left(\frac{\beta}{2\pi}\right)^3\iiint \mathfrak{z}^*(i\mathbf{f}) e^{-i\beta\mathbf{f}\cdot\boldsymbol{\xi}} \,d^3\mathbf{f}
\label{Laplacetrans2}
	.
\end{equation}
Eqs.~\eqref{Laplacetrans1} and \eqref{Laplacetrans2} are the ensemble transformation relations. 
In the thermodynamic limit and under appreciable loads \cite{neumann1985nonequivalence, neumann2013comment, manca2013response}, fluctuations become negligible and the free energies of the two ensembles are related by the Legendre transformation

\begin{equation}
	\varphi^*(\mathbf{f}) = 
	\psi^*(\boldsymbol{\xi}) - \mathbf{f}\cdot\boldsymbol{\xi}
\label{trans1}
	.
\end{equation}
Therefore, in the limit of long chains, the mechanical response of the chain can be obtained equivalently from either ensemble:

\begin{equation}
	\mathbf{f} = \frac{\partial\psi^*}{\partial\boldsymbol{\xi}}
	,\qquad
	\boldsymbol{\xi} = -\frac{\partial\varphi^*}{\partial\mathbf{f}}
	\label{endtoendgibbs}
	,
\end{equation}
and the two equilibrium distributions are related by

\begin{equation}
	\frac{P_\mathfrak{z}^\mathrm{eq}(\mathbf{f})}{P_\mathfrak{z}^\mathrm{eq}(\mathbf{f}_\mathrm{ref})} = 
	e^{-\beta(\psi_\mathrm{ref}-\varphi_\mathrm{ref}-\mathbf{f}\cdot\boldsymbol{\xi})} \left[\frac{P^\mathrm{eq}(\boldsymbol{\xi})}{P^\mathrm{eq}(\boldsymbol{\xi}_\mathrm{ref})}\right]
	.\label{trans2}
\end{equation}
Eqs.~\eqref{trans1} and \eqref{trans2} are the ensemble transformation relations in the thermodynamic limit.

%%%%%%%%%%%%%%%%%%%%%%%%%%%%%%%%%%%%%%%%%%%%%%%%%%%%%%%%%%
\subsubsection{Example polymer chain model}

In order to demonstrate the above sets of equations, we consider the EFJC model, where the polymer chain is represented by $M=N_b+1$ atoms/hinges connected in series by $N_b$ flexible links of rest length $\ell_b$ and harmonic potential stiffnesses $k_b$. 
Due to the nonzero potentials, the mechanical response of this model will be due to coupled contributions from both entropic and enthalpic effects.
The EFJC model has a Gibbs ensemble partition function that can be evaluated analytically: it is given by Fiasconaro and Falo \cite{fiasconaro2019analytical} as

\begin{equation}
	\mathfrak{z}^*(\eta) = 
	\left\{
	B_0\,
	\frac{\sinh(\eta)}{\eta}\,e^{\eta^2/2\kappa}\left[1 + \frac{\eta}{\kappa}\,\coth(\eta)\right]\right\}^{N_b}
\label{zefjc}
	,
\end{equation}
where $\eta = \beta f\ell_b$ is the non-dimensional force, $\kappa=\beta k_b\ell_b^2$ is the non-dimensional link stiffness, and $B_0=2^{5/2}\pi^{3/2}\beta\ell_b^3\kappa^{-1/2}$.
For $\kappa\to\infty$, we recover the freely joined chain (FJC) Gibbs ensemble partition function \cite{rubinstein2003polymer}.
To retrieve the Helmholtz ensemble partition function we use Eq.~\eqref{Laplacetrans2}, which in this case (spherically symmetric) as shown by Manca, et.~al. \cite{manca2012elasticity} reduces to

\begin{equation}
	\mathfrak{q}^*(\lambda) = 
	\frac{1}{2\pi^2 N_b\ell_b^3}\frac{1}{\lambda}\int_0^\infty \mathfrak{z}^*(i\eta)\sin(N_b\eta\lambda)\eta\,d\eta
\label{Laplacetrans2sym}
	,
\end{equation}
where $\lambda=\xi/N_b\ell_b$ is the chain end-to-end stretch relative to the contour length $N_b\ell_b$. 
After using Eq.~\eqref{Laplacetrans2sym} to calculate $\mathfrak{q}^*$, we then use Eq.~\eqref{psifromq} in order to calculate $\psi^*$ from $\mathfrak{q}^*$, and subsequently Eq.~\eqref{equiv1} to calculate $P^\mathrm{eq}$ from $\psi^*$. 
Although we have started from the Gibbs ensemble, we have calculated $\psi^*$ and $P^\mathrm{eq}$ exactly, which we will refer to as the Helmholtz method (Fig.~\ref{scheme1}, top pathway). 

The Helmholtz method is often computationally challenging, so simpler approximate methods are typically used. 
When the thermodynamic limit $N_b\to\infty$ is satisfied, the Legendre transformation in Eq.~\eqref{trans1} may be used to calculate the Helmholtz free energy $\psi^*$ from the Gibbs free energy $\varphi^*$.
We will refer to this method as the Gibbs-Legendre method (Fig.~\ref{scheme1}, bottom pathway). 
This method is expedient if $\varphi^*$ is known exactly, which is true in the case of the EFJC model. 
The exact value of $\varphi^*$ for the EFJC is calculated by plugging $\mathfrak{z}^*$ from Eq.~\eqref{zefjc} into Eq.~\eqref{freeenergygibbs}.
To obtain the mechanical response in order to perform the Legendre transformation, we use Eq.~\eqref{endtoendgibbs} to obtain the non-dimensional end-to-end length in the Gibbs ensemble

\begin{equation}
	\lambda(\eta) = 
	\mathcal{L}(\eta) + \frac{\eta}{\kappa}\left[1 + \frac{1 - \mathcal{L}(\eta)\coth(\eta)}{1 + (\eta/\kappa)\coth(\eta)}\right]
\label{efjcflambda}
	.
\end{equation}
Now we assume that the thermodynamic limit $N_b\to\infty$ is satisfied and use Eq.~\eqref{trans1} to calculate the Helmholtz free energy for the EFJC to be

%%%%%%%%%%%%%%%%%%%%%%%%%%%%%%%%%%%%%%%%%%%%%%%%%%%%%%%%%%
\begin{figure*}[t]
\begin{center}
$
\begin{array}{ccc}
\includegraphics{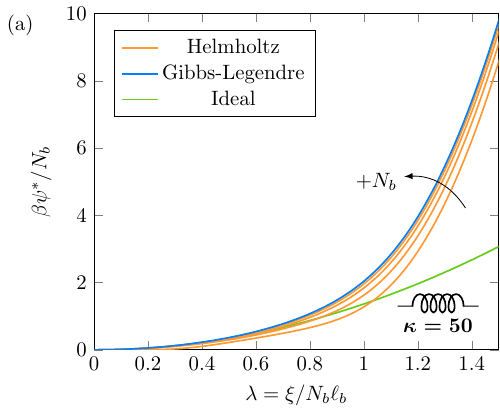}
& \qquad &
\includegraphics{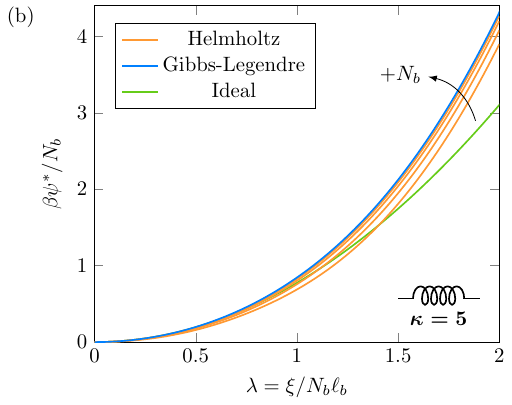}
\end{array}
$
\vspace{-5pt}
\end{center}
\caption{\label{fig1}
	Non-dimensional free energy per link versus end-to-end stretch for $N_b = 3$, 5, 10, and 25, for (a) $\kappa=50$ and (b) $\kappa = 5$. 
  The free energy is plotted using both the Helmholtz and Gibbs-Legendre methods, as well as using the ideal model valid for $\lambda\ll 1$; the Gibbs-Legendre and ideal results are independent of $N_b$. 
}
\end{figure*}
%%%%%%%%%%%%%%%%%%%%%%%%%%%%%%%%%%%%%%%%%%%%%%%%%%%%%%%%%%

\begin{eqnarray}
	\psi^*(\lambda) = &&
	N_bkT\left\{
	\eta\mathcal{L}(\eta) + \ln\left[\frac{\eta}{\sinh(\eta)}\right]
	\right. \nonumber \\ && \left. 
	- \ln\left[1 + \frac{\eta}{\kappa}\,\coth(\eta)\right] - \ln B_0
	\right. \nonumber \\ && \left. 
	+ \frac{\eta^2}{\kappa}\left[\frac{1}{2} + \frac{1 - \mathcal{L}(\eta)\coth(\eta)}{1 + (\eta/\kappa)\coth(\eta)}\right]
	\right\}
\label{psiefjclim}
	,
\end{eqnarray}
where we solve for $\eta=\eta(\lambda)$ using Eq.~\eqref{efjcflambda} in order to get $\psi^*=\psi^*(\lambda)$. 
The equilibrium distribution in the thermodynamic limit is then calculated using Eq.~\eqref{equiv1},

\begin{eqnarray}
	P^\mathrm{eq}(\lambda) &&= 
	\frac{1}{\ell_b^3C}
	\left(
	\frac{  
	\sinh(\eta)\left[1 + (\eta/\kappa)\coth(\eta)\right]
	}{
	\eta\exp\left[\eta\mathcal{L}(\eta)\right]
	}
	\right. \nonumber \\ 
	&&\left. \times
	\exp\left\{\frac{\eta^2}{\kappa}\left[\frac{1}{2} + \frac{1 - \mathcal{L}(\eta)\coth(\eta)}{1 + (\eta/\kappa)\coth(\eta)}\right]\right\}
	\right)^{N_b}
\label{peqefjclim}
	,~~~~
\end{eqnarray}
where $C=C(N_b,\kappa)$ is such that the distribution is normalized.
For $\kappa\to\infty$ in Eqs.~\eqref{efjcflambda} and \eqref{peqefjclim}, we recover the FJC mechanical response $\lambda=\mathcal{L}(\eta)$ and probability distribution in the thermodynamic limit \cite{treloar1949physics}. 

An alternative approximation method is to assume a Gaussian distribution for the equilibrium distribution.
This assumption is valid in the limit $N_b\to\infty$ due to the central limit theorem. 
In order to determine this Gaussian distribution for the EFJC model, we first approximate the mechanical response in Eq.~\eqref{efjcflambda} for small forces $(\eta\ll 1)$ by the linear relation

\begin{equation}
	\lambda(\eta) = 
	\frac{\eta}{c_\kappa}
	,\qquad
	c_\kappa =
	\frac{\kappa(\kappa+1)}{\kappa^2+6\kappa+3}
	,
\end{equation}
and subsequently the free energy in Eq.~\eqref{psiefjclim} by a quadratic relation for $\eta\ll 1$.
Combining these results yields the small stretch $(\lambda\ll 1)$ free energy

\begin{equation}
	\psi^*(\lambda) = 
	\frac{3}{2}\,c_\kappa N_bkT\lambda^2
\label{psigaussK}
	.
\end{equation} 
We now make use of this small stretch approximation to construct the equilibrium distribution for $N_b\to\infty$ using Eq.~\eqref{equiv1}, which is then

\begin{equation}
	P^\mathrm{eq}(\lambda) = 
	\left(\frac{3c_\kappa}{2\pi N_b\ell_b^2}\right)^{3/2} \exp\left(-\frac{3}{2}\,c_\kappa N_b \lambda^2\right)
	.
\label{PgaussK}
\end{equation} 
We remark that this distribution is a valid approximation for any stretch as long as $N_b\to\infty$, so it is common to utilize this equilibrium distribution with the Gibbs-Legendre method free energy function in order to approximate the full Helmholtz method. 

%%%%%%%%%%%%%%%%%%%%%%%%%%%%%%%%%%%%%%%%%%%%%%%%%%%%%%%%%%
\begin{figure*}[t]
\begin{center}
$
\begin{array}{ccc}
\includegraphics{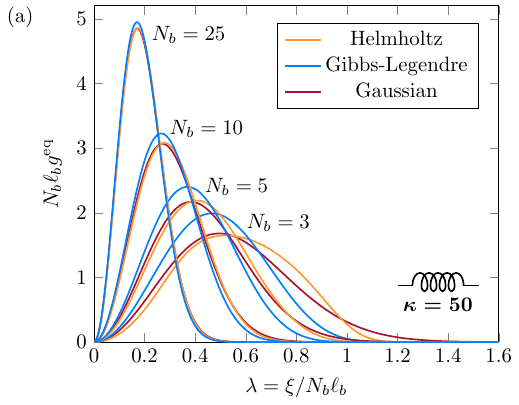}
& \qquad &
\includegraphics{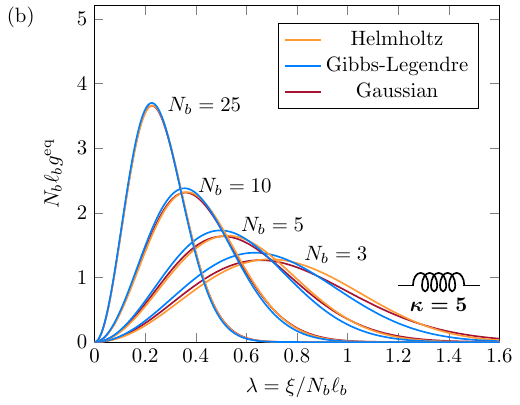}
\end{array}
$
\vspace{-5pt}
\end{center}
\caption{\label{fig2}
	Non-dimensional equilibrium radial distribution function versus end-to-end stretch for $N_b = 3$, 5, 10, and 25, for (a) $\kappa=50$ and (b) $\kappa = 5$. 
  The distribution is plotted using the Helmholtz, Gibbs-Legendre, and Gaussian methods. 
}
\end{figure*}
%%%%%%%%%%%%%%%%%%%%%%%%%%%%%%%%%%%%%%%%%%%%%%%%%%%%%%%%%%

Since the Gibbs-Legendre method is often used to approximate the true Helmholtz free energy, we plot the EFJC non-dimensional free energy $(\beta\psi^*/N_b)$ as a function of end-to-end chain stretch $(\lambda)$ for $\kappa=50$ and varying $N_b$ in Fig.~\ref{fig1}(a), obtained using both the Helmholtz and the Gibbs-Legendre methods, as well as the ideal chain free energy.
See that for small values of $N_b$ the difference in free energy between the Helmholtz and Gibbs-Legendre methods is quite considerable, such as for $N_b=3$ where the relative difference is nearly constant at 60\% for $\lambda\in(0.5,1)$.
As $N_b$ increases, the difference between the two methods shrinks, becoming quite small when $N_b=25$.
We can also observe that the ideal chain free energy, which matches the Gibbs-Legendre method free energy at small stretch, does not match that of the Helmholtz method until $N_b$ becomes large.
We repeat this analysis for the smaller EFJC link stiffness $(\kappa=5)$ in Fig.~\ref{fig1}(b), where we observe the same trends but overall smaller differences among the methods. 
This can be understood by reconsidering the Gibbs ensemble partition function in Eq.~\eqref{zefjc} and the ensemble transformation relation in Eq.~\eqref{Laplacetrans2sym}. 
First consider the case of $N_b\to\infty$, where we will receive $\mathfrak{q}^*(\lambda)\to\mathfrak{z}^*(\eta)e^{-N_b\eta\lambda}$ from Eq.~\eqref{Laplacetrans2sym}, and where the Gibbs-Legendre method results will exactly match that of the Helmholtz method. 
This is because $\mathfrak{z}^*(i\eta)$ will decay rapidly as a function of $\eta$ when $N_b$ becomes large and effectively contain a Dirac delta function.
Now for $\kappa\to 0$, we see that $\mathfrak{z}^*(i\eta)$ will also decay rapidly as a function of $\eta$, which will also act as a Dirac delta function via one definition,

\begin{equation}
	\delta(\eta) = 
	\lim_{\kappa\to 0^+} \frac{1}{\sqrt{2\pi\kappa}}\,e^{-\eta^2/2\kappa}
	,
\end{equation}
which appears in $\mathfrak{z}^*(i\eta)$ after recalling that $B_0\propto 1/\sqrt{\kappa}$ in Eq.~\eqref{zefjc}. 
This is why we observe smaller differences between the Gibbs-Legendre and Helmholtz methods as $\kappa$ decreases.
It can also be understood intuitively as a decreasing correlation between the links: the link degrees of freedom in the Gibbs ensemble are completely independent, while that in the Helmholtz ensemble are because of the end-to-end length constraint. 
As $\kappa$ decreases the link degrees of freedom in the Helmholtz ensemble become increasingly independent of each other, approaching the $\kappa=0$ limit where they are completely independent.

We plot the EFJC non-dimensional radial distribution function $(N_b\ell_b g^\mathrm{eq})$ as a function of end-to-end stretch for $\kappa=50$ and varying $N_b$ in Fig.~\ref{fig2}(a).
The radial distribution function is plotted using both the Helmholtz and the Gibbs-Legendre methods, as well as the $N_b\to\infty$ limiting Gaussian distribution.
The Gibbs-Legendre distribution tends to be quite different from the Helmholtz distribution for small values of $N_b$, while the Gaussian distribution tends to be a bit closer.
By $N_b=25$, the Gaussian and Helmholtz distributions become nearly indistinguishable, and the Gibbs-Legendre distribution retains only a small difference from the other two.
We repeat this analysis for a smaller EFJC link stiffness ($\kappa=5$) in Fig.~\ref{fig2}(b), where we observe the same trends but overall smaller differences among the methods. 
This difference is again explained by the more rapidly decaying $\mathfrak{z}^*(i\eta)$ in Eq.~\eqref{Laplacetrans2sym} as $\kappa$ decreases, as previously discussed. 
Though the Gibbs-Legendre method free energy is immensely closer to the Helmholtz method free energy than the ideal chain free energy, we see here that the Gaussian distribution -- obtained from the ideal chain free energy using the distribution-behavior correspondence in Eq.~\eqref{equiv1} --  tends to be much closer to the Helmholtz method distribution than the Gibbs-Legendre method distribution. 
Looking back to Fig.~\ref{fig1}, this is likely because the Gibbs-Legendre method overestimates the single chain free energy increase with stretch for smaller $N_b$, resulting in an underestimate of the probability of chains at larger stretch observed in Fig.~\ref{fig2} due to the distribution-behavior correspondence relations. 

%%%%%%%%%%%%%%%%%%%%%%%%%%%%%%%%%%%%%%%%%%%%%%%%%%%%%%%%%%
\subsubsection{\label{sec2.1.3}Distribution evolution}

We introduce $P(\boldsymbol{\xi},t)$ as the probability density distribution of chains with end-to-end vector $\boldsymbol{\xi}$ at time $t$, which we presume to initially be in the equilibrium distribution, $P(\boldsymbol{\xi},0)=P^\mathrm{eq}(\boldsymbol{\xi})$. 
Liouville's equation \cite{mcq} describes the evolution of a probability density $P$ in the single chain phase space (atomic coordinates and momenta) as

\begin{equation}
	\frac{\partial P}{\partial t} = 
	-\sum_{j=1}^{M} \left(\frac{\partial P}{\partial \mathbf{q}_j} \cdot \dot{\mathbf{q}}_j + \frac{\partial P}{\partial \mathbf{p}_j} \cdot \dot{\mathbf{p}}_j\right)
	.
\end{equation}
When we apply Liouville's equation to $P(\boldsymbol{\xi},t)$, the only nonzero derivative we retrieve is that relating to the chain end-to-end vector $\boldsymbol{\xi}=\mathbf{q}_{M}-\mathbf{q}_1$, and thus the evolution law for the distribution of end-to-end vectors is

\begin{equation}\label{nevopre}
	\frac{\partial P}{\partial t} = 
	-\frac{\partial P}{\partial \boldsymbol{\xi}} \cdot \dot{\boldsymbol{\xi}}
	,
\end{equation}
where $\dot{\boldsymbol{\xi}} = \dot{\boldsymbol{\xi}}(\boldsymbol{\xi},t)$ is left to be prescribed.

%%%%%%%%%%%%%%%%%%%%%%%%%%%%%%%%%%%%%%%%%%%%%%%%%%%%%%%%%%
\subsection{\label{sec2.2}Macroscopic constitutive theory}

In order to extend our theory into the macroscale, we prescribe an affine deformation to an incompressible network and analytically solve for the distribution evolution.
Equipped with this connection between the statistical and continuum mechanics of the polymer network, we use the Coleman-Noll procedure \cite{coleman13noll} to develop the macroscopic constitutive theory. 
We choose the deformation gradient $\mathbf{F}$ and the temperature $T$ as the independent thermodynamic state variables.
We presume these thermodynamic state variables to be complete, allowing us to consider time derivatives of constitutive functions to be implicit, i.e. we may expand them in terms of the time derivatives of the state variables. 
After some derivation -- including the full treatment of a boundary integral term that, until now, has been either omitted or otherwise assumed to be zero -- we ultimately retrieve a closed-form relation for the Cauchy stress in terms of the applied deformation, the chain free energy function, and the equilibrium distribution of end-to-end vectors.

\subsubsection{Macroscopic connection}

We assume that the evolution of end-to-end vectors is affine with the deformation, $\dot{\boldsymbol{\xi}} = \mathbf{L}\cdot\boldsymbol{\xi}$, where $\mathbf{L}=\dot{\mathbf{F}}\cdot\mathbf{F}^{-1}$ is the velocity gradient. Eq.~\eqref{nevopre} then becomes

\begin{equation}\label{nevo}
	\frac{\partial {P}}{\partial t} =
	-\left(\frac{\partial P}{\partial\boldsymbol{\xi}}\right)\cdot\mathbf{L}\cdot\boldsymbol{\xi}
	.
\end{equation}
This first order linear partial differential equation can be solved analytically using the method of characteristics (See Appendix~\ref{appPDE}). Under the initial conditions $\mathbf{F}(0)=\mathbf{1}$ and ${P}(\boldsymbol{\xi},0) = {P}^\mathrm{eq}(\boldsymbol{\xi})$, the solution is 

\begin{equation}\label{nsol}
	{P}(\boldsymbol{\xi},t) = 
	{P}^\mathrm{eq}\left[\mathbf{F}^{-1}(t)\cdot\boldsymbol{\xi}\right]
	,
\end{equation}
which simply states that the probability density of a chain having end-to-end vector $\boldsymbol{\xi}$ at time $t$ is equal to the probability density of that end-to-end vector mapped backward to the corresponding end-to-end vector in the equilibrium distribution.

\subsubsection{Second law analysis}

Now that we are equipped with the probability distribution of polymer chains within the network as a function of the deformation, we write the current Helmholtz free energy density of the network ($a$) by integrating the probability-weighted free energy function over all end-to-end vectors,

\begin{equation}
	a(t) = 
	n\iiint P(\boldsymbol{\xi},t) \psi^*(\boldsymbol{\xi}) \,d^3\boldsymbol{\xi} - p(J-1)
	,
\label{acon}
\end{equation}
where $n=N/V$ is the number density of chains and $p$ is the pressure enforcing the incompressibility constraint that $J=\det(\mathbf{F})=1$.
Note that we have only included contributions related to the chain configuration integral and left out those related to the chain momentum integral.
This is because the latter terms will only introduce spherical terms to the stress (ideal gas law) and therefore can be lumped into the pressure without loss of generality.
Thermodynamically admissible processes satisfy the Clausius-Duhem inequality \cite{truesdell2004non},

\begin{equation}\label{CDI}
	\dot{a} + s\dot{T} - \boldsymbol{\sigma}:\mathbf{L} \leq 0
	,
\end{equation}
where $s$ is the entropy density and $\boldsymbol{\sigma}$ is the Cauchy stress tensor. 
This reduced form of the Clausius-Duhem inequality involves several classical assumptions that are standard in the Coleman-Noll procedure, such as the neglect of non-mechanical work, the constitutive relations for the entropy flux and entropy source, and in this case Fourier's law for the heat flux \cite{paolucci2016continuum}.
We expand the implicit time derivative of the Helmholtz free energy density using our complete set of state variables,

\begin{eqnarray}
	\dot{a} = &&
	\left(\frac{\partial a}{\partial t}\right)_T + \left(\frac{\partial a}{\partial t}\right)_\mathbf{F}
	\\ = &&
	\left(\frac{\partial a}{\partial \mathbf{F}}\right)_T:\dot{\mathbf{F}} +\left(\frac{\partial a}{\partial T}\right)_\mathbf{F}\dot{T}
	,
\end{eqnarray}
and substitute this result back into Eq.~\eqref{CDI} for

\begin{equation}
\left[\left(\frac{\partial a}{\partial T}\right)_\mathbf{F} + s\right]\dot{T} + \left[\left(\frac{\partial a}{\partial \mathbf{F}}\right)_T\cdot\mathbf{F}^T - \boldsymbol{\sigma}\right]:\mathbf{L} \leq 0
.
\end{equation}
We now consider the set of processes where the deformation is held fixed, $\mathbf{L}=\mathbf{0}$, and the temperature is varied arbitrarily. 
Since $\dot{T}$ can be any real number, positive or negative, and this inequality must hold, we see that the term in the first set of brackets must always be zero and we  receive the expected constitutive relation for the entropy density 

\begin{equation}
	s =
	-\left(\frac{\partial a}{\partial T}\right)_\mathbf{F}
	,
\end{equation}
and after going back to our original derivative notation, we are left with the remaining dissipation inequality

\begin{equation}
	\left(\frac{\partial a}{\partial t}\right)_T  - \boldsymbol{\sigma}:\mathbf{L}
	\leq 0
	.
\label{stressineq}
\end{equation}
Several steps are then taken in order to proceed from Eq.~\eqref{stressineq} and retrieve the stress.
We first neglect dissipative stresses, thus taking the equality in Eq.~\eqref{stressineq} and receiving a hyperelastic stress.
Next, we take Eq.~\eqref{acon} and assume spherical symmetry in $\psi^*$, which causes the stress to be non-polar. 
We then require that $\psi^*$ grows sufficiently fast as $\xi\to\infty$, in order to show that the boundary integral term resulting from integration by parts is zero.
The full derivation is presented in Appendix~\ref{appstress} and yields the stress to be

\begin{eqnarray}
	\boldsymbol{\sigma}(t) =&&
	n\iiint {P}^\mathrm{eq}\left[\mathbf{F}^{-1}(t)\cdot\boldsymbol{\xi}\right] \left(\frac{\partial\psi^*}{\partial\xi}\right) \left(\frac{\boldsymbol{\xi}\boldsymbol{\xi}}{\xi}\right) \,d^3\boldsymbol{\xi} 
	\nonumber \\ 
	&&- \left[p^\mathrm{eq} + \Delta p(t)\right]\mathbf{1}
\label{stressmixedrep},
\end{eqnarray}
where $\mathbf{1}$ is the identity tensor, the differential pressure $\Delta p(t)$ enforces incompressibility, and the equilibrium pressure $p^\mathrm{eq}$ (from $\boldsymbol{\sigma}(0)=\mathbf{0}$) is

\begin{equation}\label{peqeqn}
	p^\mathrm{eq} = 
	\frac{n}{3}
	\int g^\mathrm{eq}(\xi)\left(\frac{\partial\psi^*}{\partial\xi}\right)\,\xi\,d\xi 
	.
\end{equation}
The derivative of the chain Helmholtz free energy can be replaced with the force using Eq.~\eqref{endtoendgibbs}, but one must be careful to ensure that the force is computed in the Helmholtz ensemble: the force from the Gibbs ensemble may only be used in the thermodynamic limit ($N_b\to\infty$).
If we utilize the ideal chain free energy from Eq.~\eqref{psigaussK}, one can easily show that the Neo-Hookean model results, as expected (see Appendix~\ref{neohook}). 

%%%%%%%%%%%%%%%%%%%%%%%%%%%%%%%%%%%%%%%%%%%%%%%%%%%%%%%%%%
\subsection{\label{sec2.3}Implementation}

To close this section, we would like to point out some important aspects of the model implementation. 
It happens that Eq.~\eqref{Laplacetrans2sym} is difficult to evaluate with the EFJC partition function in Eq.~\eqref{zefjc} for moderate to large $N_b$, which is due to the integrand oscillating rapidly and decaying slowly. 
While certain integration schemes may perform reasonably well for small or large $N_b$, it is most desirable to use an integration scheme that remains accurate for the full range of $N_b$ being considered.
To evaluate this integral with high precision, we used the double exponential quadrature scheme presented by Ooura and Mori \cite{ooura1991double} and their Fortran script \texttt{intdeo.f} that implements it, as well as the arbitrary precision Fortran package \texttt{MPFUN2015} provided by Bailey \cite{bailey2015mpfun2015}.
These calculations were carried out using the Extreme Science and Engineering Discovery Environment (XSEDE) \texttt{Stampede2} cluster \cite{towns2014xsede}.
When calculating the integrals in Eq.~\eqref{stressmixedrep} in order to retrieve the stress, it is unwieldy to repeatedly call a function to exactly evaluate $\psi^*(\lambda)$ for the EFJC as $\lambda\to\infty$.
See from Eqs.~\eqref{efjcflambda}--\eqref{psiefjclim} that $\lambda\sim\eta/\kappa$ and $\beta\psi^*\sim N_b\eta^2/\kappa$ as $\eta\to\infty$, which combined shows that $\beta\psi^*\sim N_b\kappa\lambda^2$ as $\lambda\to\infty$.
The neglected terms in this asymptotic relation for $\psi^*(\lambda)$ are quite small, so the relation is accurate even for $\lambda$ only moderately above unity. 
Therefore, in order to greatly speed up the computation of the stress at negligible cost to accuracy, we fit a quadratic function to $\psi^*$ for large $\lambda$ and call this function instead when $\lambda$ is above a certain value. 

%%%%%%%%%%%%%%%%%%%%%%%%%%%%%%%%%%%%%%%%%%%%%%%%%%%%%%%%%%
\section{\label{sec3}Macroscopic Results}

Now that we have fully formulated the theory, we are able to explicitly examine the effects that changes in the statistical description have on the macroscopic mechanics. 
Traditionally in polymer network constitutive modeling, the Helmholtz ensemble has been approximated using the Legendre transformation from the Gibbs ensemble, so we will start by examining the difference in macroscopic mechanical response when using these Helmholtz and Gibbs-Legendre methods. 
Another common choice in these constitutive models is to assume that the equilibrium distribution is Gaussian, so we will then examine how the true Helmholtz ensemble mechanical response differs from that assuming the Gaussian distribution and the Gibbs-Legendre free energy function; we will refer to this as the Gibbs-Legendre-Gaussian method. 
In both of these studies, we will see that a long enough polymer chain causes all approaches to result in the same mechanical response.
This convergence occurs before the $N_b\to\infty$ limit represented by choosing the ideal chain free energy function and the Gaussian distribution, which is the Neo-Hookean model.

%%%%%%%%%%%%%%%%%%%%%%%%%%%%%%%%%%%%%%%%%%%%%%%%%%%%%%%%%%
\begin{figure*}[t!]
\begin{center}
$
\begin{array}{ccc}
\includegraphics{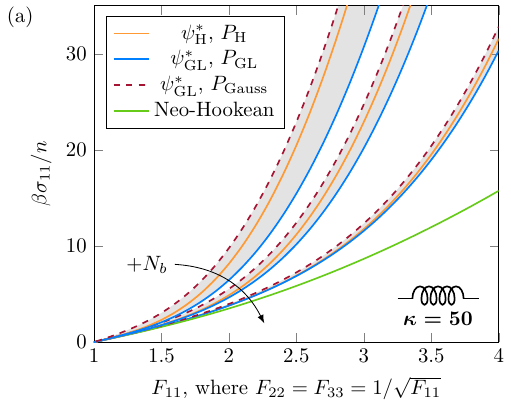}
& \qquad &
\includegraphics{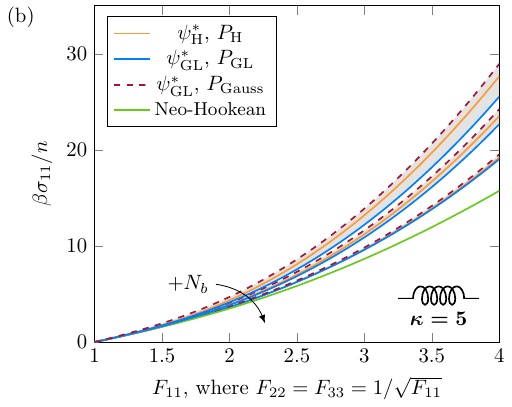}
\end{array}
$
\vspace{-5pt}
\end{center}
\caption{\label{fig3}
	Non-dimensional uniaxial stress-stretch results for the EFJC network with (a) $\kappa=50$ and (b) $\kappa=5$, for $N_b=5$, 10, and 25.
  This mechanical response is plotted using the true method (Helmholtz) and two approximation methods (Gibbs-Legendre, Gibbs-Legendre-Gaussian).
  Shading indicates equal $N_b$ value.
  The Neo-Hookean response is included as reference.
}
\end{figure*}
%%%%%%%%%%%%%%%%%%%%%%%%%%%%%%%%%%%%%%%%%%%%%%%%%%%%%%%%%%

\begin{figure*}[t!]
\begin{center}
$
\begin{array}{ccc}
\includegraphics{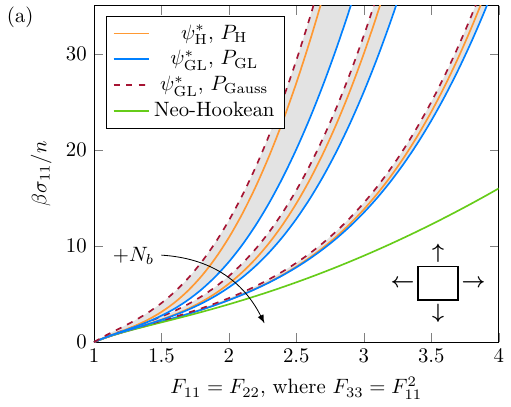}
& \qquad &
\includegraphics{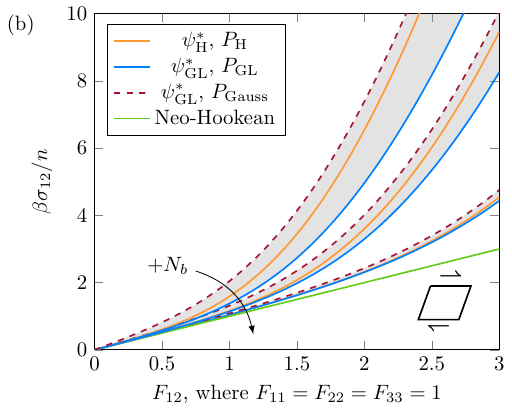}
\end{array}
$
\vspace{-5pt}
\end{center}
\caption{\label{fig4}
  Non-dimensional (a) equibiaxial and (b) simple shear stress-stretch results for the EFJC network with $\kappa=50$ for $N_b=5$, 10, and 25.
  This mechanical response is plotted using the true method (Helmholtz) and two approximation methods (Gibbs-Legendre, Gibbs-Legendre-Gaussian).
  Shading indicates equal $N_b$ value.
  The Neo-Hookean response is included as reference.
}
\end{figure*}

We apply the Helmholtz and Gibbs-Legendre methods to a polymer network modeled to consist of EFJCs.
$\mathfrak{z}^*$ for the EFJC is given by Eq.~\eqref{zefjc}.
The Helmholtz method takes $\mathfrak{z}^*$ and uses Eq.~\eqref{Laplacetrans2sym} to compute $\mathfrak{q}^*$, then computes $\psi^*$ using Eq.~\eqref{psifromq} and $P^\mathrm{eq}$ using Eq.~\eqref{trans1}.
The Gibbs-Legendre method assumes $N_b\to\infty$ in order to use $\psi^*$ in Eq.~\eqref{psiefjclim} and $P^\mathrm{eq}$ in Eq.~\eqref{peqefjclim}.
In both methods, we compute the stress under uniaxial tension in the 1-direction using Eq.~\eqref{stressmixedrep}, where the deformation gradient is diagonal with components due to symmetry and incompressibility, $F_{22}=F_{33}=1/\sqrt{F_{11}}$. 
In Fig.~\ref{fig3}(a) we plot the non-dimensional uniaxial stress $\beta\sigma_{11}/n$ versus the applied stretch $F_{11}$ using the non-dimensional EFJC stiffness $\kappa=50$ and an increasing numbers of links $N_b=5$, 10, and 25.
The Neo-Hookean stress-stretch response -- retrieved through using Eq.~\eqref{psigaussK} for $\psi^*$ and Eq.~\eqref{PgaussK} for $P^\mathrm{eq}$ -- is included for reference. 
For small numbers of links such as $N_b=5$, the Gibbs-Legendre method drastically underestimates the overall stiffness of the true stress response from the Helmholtz method.
This difference shrinks as $N_b$ increases, becoming only a tiny (but increasing with stretch) difference when $N_b=25$, analogous to the differences between the free energy functions and the equilibrium distributions shrinking in Figs.~\ref{fig1} and \ref{fig2}.
We see that the two methods seem to converge before the limit $N_b\to\infty$ is truly reached, where the Neo-Hookean mechanical response would be retrieved. 
We have repeated the same analysis for the lower stiffness ($\kappa=5$) in Fig.~\ref{fig3}(b), where we observe the same behavior as $N_b$ increases, but in general less difference between the two methods compared to the $\kappa=50$ case for any $N_b$. 
These features are independent of loading mode, as is evident by the above analyses implemented for equibiaxial tension and simple shear (Fig.~\ref{fig4}).
For equibiaxial tension we apply $F_{11}=F_{22}$, where incompressibility requires $F_{33}=1/F_{11}^2$, and for simple shear we apply $F_{12}$, where $F_{11}=F_{22}=F_{33}=1$. 

As previously mentioned, the equilibrium distribution $P^\mathrm{eq}$ tends towards Gaussian as $N_b\to\infty$.
It is common for polymer network constitutive models to choose a non-ideal free energy function $\psi^*$ but assume $N_b$ is large enough to warrant the use of the Gaussian $P^\mathrm{eq}$, rather than the $P^\mathrm{eq}$ from the distribution-behavior correspondence relation given by Eq.~\eqref{equiv2}.
For the case of the EFJC, the limit $N_b\to\infty$ results in the Gaussian $P^\mathrm{eq}$ given by Eq.~\eqref{PgaussK}.
In attempting to approximate the true Helmholtz method in the limit as $N_b\to\infty$, one would then assume a Gaussian equilibrium distribution and either use the Helmholtz method or Gibbs-Legendre method for $\psi^*$.
We will neglect the case of the Helmholtz method $\psi^*$ combined with the Gaussian $P^\mathrm{eq}$, since by distribution-behavior correspondence, one could simply find the true $P^\mathrm{eq}$ after knowing $\psi^*$.
In either case, the macroscopic mechanical response will converge to that of the true Helmholtz method when $N_b$ becomes sufficiently large.

We apply the Gibbs-Legendre-Gaussian method to a polymer network modeled to consist of EFJCs.
The Gibbs-Legendre-Gaussian method uses $\psi^*$ from Eq.~\eqref{psiefjclim} and the Gaussian $P^\mathrm{eq}$ from Eq.~\eqref{PgaussK}.
Taking the non-dimensional EFJC stiffness $\kappa=50$, in Fig.~\ref{fig3}(a) we plot the non-dimensional stress $\beta\sigma_{11}/n$ versus the applied stretch $F_{11}$ for increasing numbers of links $N_b=5$, 10, and 25.
The Gibbs-Legendre-Gaussian method does well matching the tangent stiffness and keeping the relative error small at larger stretches, but tends to do poorly at small to intermediate stretches ($F_{11}\leq 2$).
We also see that this method seems to converge to the Helmholtz method stress-stretch response before the limit $N_b\to\infty$ is truly reached, where the Neo-Hookean mechanical response would be retrieved. 
We have repeated the same analysis for the lower stiffness $\kappa=5$ in Fig.~\ref{fig3}(b), where we observe the same trends but smaller relative errors, since as previously discussed and shown in Fig.~\ref{fig2}, the distributions from different methods become more alike for smaller $\kappa$.

When comparing the three methods in Figs.~\ref{fig3} and \ref{fig4}, we see that the Gibbs-Legendre method tends to underestimate the true mechanical response given by the Helmholtz method, and the Gibbs-Legendre-Gaussian method tends to overestimate it, especially at small stretches. 
To understand this further, we can consider the initial moduli in each case by applying an infinitesimal deformation $\mathbf{F}=\mathbf{1}+\mathbf{E}$, where $\mathbf{E}$ is the infinitesimal strain tensor. 
Straightforward analysis shows (see Appendix~\ref{appsmall}) that the stress from Eq.~\eqref{stressmixedrep} then becomes

\begin{equation}
	\boldsymbol{\sigma}(t) =
	\mu\mathbf{E}(t) - \Delta p(t)\mathbf{1}
	,
\label{stress_small}
\end{equation}
where method-specific shear modulus $\mu$ is given by

%%%%%%%%%%%%%%%%%%%%%%%%%%%%%%%%%%%%%%%%%%%%%%%%%%%%%%%%%%
%%%%%%%%%%%%%%%%%%%%%%%%%%%%%%%%%%%%%%%%%%%%%%%%%%%%%%%%%%

\begin{equation}
	\mu = 
	\frac{8\pi}{15}\,nkT\iiint \left(-\frac{\partial{P}^\mathrm{eq}}{\partial\xi}\right)\left(\frac{\partial\beta\psi^*}{\partial\xi}\right)\xi^4\,d\xi
	.
\label{modulus_small}
\end{equation}
We find $\Delta p$ by enforcing incompressibility via $\mathrm{tr}(\mathbf{E})=0$, from which we find the initial moduli to be $3\mu/2$, $3\mu$, and $\mu$ for uniaxial tension, equibiaxial tension, and simple shear, respectively. 
For the Neo-Hookean model we receive $\mu=2nkT$, as expected, whereas for the other methods we cannot analytically compute the integral but in general receive $\mu\geq 2nkT$.
We can, however, compute the shear modulus in specific cases: for $N_b=5$ and $\kappa=50$, $\mu=2.2322$ for the Helmholtz method, $\mu=2.1914$ for the Gibbs-Legendre method, and $\mu=3.0215$ for the Gibbs-Legendre-Gaussian method. 
See that while the Gibbs-Legendre method underestimates the modulus, the Gibbs-Legendre-Gaussian method drastically overestimates it, which is why we observe a poor performance of the Gibbs-Legendre-Gaussian method at small stretches in Figs.~\ref{fig3} and \ref{fig4}. 
We see this difference even persists when $N_b=25$ and $\kappa=50$ -- where $\mu=2.0309$, 2.0306, and 2.1113 -- and  when $N_b=5$ and $\kappa=5$ -- where $\mu=2.1020$, 2.0804, and 2.3381 -- for the Helmholtz, Gibbs-Legendre, and Gibbs-Legendre-Gaussian methods, respectively.
These differences in modulus occur because the Gibbs-Legendre-Gaussian method ignores distribution-behavior correspondence.
The Gibbs-Legendre method tends to overestimate the free energy (Fig.~\ref{fig1}), which causes the Gibbs-Legendre method distribution (Fig.~\ref{fig2}) to underestimate the equilibrium amount of chains at larger stretch.
These two inaccuracies then naturally cancel to some extent when integrating for the modulus, but when a non-corresponding distribution is instead used -- such as in the Gibbs-Legendre-Gaussian method -- this cancellation does not occur. 
The Gaussian distribution (Fig.~\ref{fig2}) predicts an increased equilibrium number of chains at larger stretch, which combined with the Gibbs-Legendre method overestimated free energy will produce a significantly overestimated modulus. 
This also explains the poor convergence of the Gibbs-Legendre-Gaussian method as $N_b$ increases, which is especially evident from Fig.~\ref{fig4}(b) where the Gibbs-Legendre method is nearly exact for $N_b=25$ while the Gibbs-Legendre-Gaussian method is not.
Therefore, it seems that it is better to obey distribution-behavior correspondence in using the Gibbs-Legendre method for small to intermediate stretches and/or larger number of links.
For a large stretch and small number of links, it is seemingly better to instead use the method with the more accurate equilibrium distribution, which here is the Gibbs-Legendre-Gaussian method.
For a truly large number of links, we can also be certain that either method will produce accurate approximations of the Helmholtz method mechanical response. 

%%%%%%%%%%%%%%%%%%%%%%%%%%%%%%%%%%%%%%%%%%%%%%%%%%%%%%%%%%
%%%%%%%%%%%%%%%%%%%%%%%%%%%%%%%%%%%%%%%%%%%%%%%%%%%%%%%%%%
\section{Conclusion}

We have performed a fundamental statistical mechanical derivation in order to account for the naturally occurring correspondences between the mechanical behavior of a single polymer chain and the equilibrium distribution of a network of such chains. 
Correspondences between different single-chain thermodynamic ensembles -- both exact and in the thermodynamic limit -- as well as the Gaussian limit of the equilibrium distribution in either ensemble, were also accounted for and discussed in detail using the extensible freely jointed chain model as an example.
This elaborate framework was then kept in-tact as we considered the macroscopic constitutive theory of the polymer network and derived the Cauchy stress in terms of the affine deformation of a general network of polymer chains. 
We used this constitutive relation for the stress to illustrate that important distinctions in the statistical description persist to play an important role in the observed macroscopic mechanical response, at least until the number of links in the polymer chains becomes large.
Obeying the distribution-behavior correspondence relations allowed a more accurate approximation of the macroscopic stress at small to intermediate deformations and/or longer chain lengths, even though this corresponding equilibrium distribution was a worse match to the true distribution than the Gaussian approximation. 
However when chains are short and the deformation is large, we saw that it was better to utilize the Gaussian distribution, which can be attributed to the extensive evolution the initial distribution undergoes during large deformations. 
This meticulous treatment is vital for future constitutive model construction and is readily applicable to more complex polymer systems.
This macroscopic framework is readily compatible with any chain models that allow the equilibrium probability distribution to be normalized and are infinitely extensible. 
Critically, this includes common biopolymer models such as the wormlike chain model as long as the extensible forms are used.
This framework does not accommodate potentials that simulate bond breaking such as the Morse potential, but bond breaking could be captured by instead including a reaction pathway to broken chains.  

%%%%%%%%%%%%%%%%%%%%%%%%%%%%%%%%%%%%%%%%%%%%%%%%%%%%%%%%%%
%%%%%%%%%%%%%%%%%%%%%%%%%%%%%%%%%%%%%%%%%%%%%%%%%%%%%%%%%%
\begin{acknowledgments}
	This material is based in part upon work supported by the National Science Foundation under Grant No. CAREER-1653059.
	Calculations in this work used the Extreme Science and Engineering Discovery Environment (XSEDE) \texttt{Stampede2} cluster, which is supported by National Science Foundation Grant ACI-1548562.
\end{acknowledgments}
%%%%%%%%%%%%%%%%%%%%%%%%%%%%%%%%%%%%%%%%%%%%%%%%%%%%%%%%%%
%%%%%%%%%%%%%%%%%%%%%%%%%%%%%%%%%%%%%%%%%%%%%%%%%%%%%%%%%%

%%%%%%%%%%%%%%%%%%%%%%%%%%%%%%%%%%%%%%%%%%%%%%%%%%%%%%%%%%
%%%%%%%%%%%%%%%%%%%%%%%%%%%%%%%%%%%%%%%%%%%%%%%%%%%%%%%%%%
\appendix
%%%%%%%%%%%%%%%%%%%%%%%%%%%%%%%%%%%%%%%%%%%%%%%%%%%%%%%%%%
%%%%%%%%%%%%%%%%%%%%%%%%%%%%%%%%%%%%%%%%%%%%%%%%%%%%%%%%%%

%%%%%%%%%%%%%%%%%%%%%%%%%%%%%%%%%%%%%%%%%%%%%%%%%%%%%%%%%%
\section{\label{appPDE}Solution for the network distribution}

We seek to analytically solve Eq.~\eqref{nevo} in order to evaluate the probability distribution $P(\boldsymbol{\xi},t)$ at any time under the deformation $\mathbf{F}(t)$. 
A new set of variables is taken: $\boldsymbol{\alpha}=(\boldsymbol{\xi},t)^T$ with gradient $\nabla_{\boldsymbol{\alpha}}=(\partial/\partial\boldsymbol{\xi},\partial/\partial t)^T$ and vector $\mathbf{b}=(\left[\mathbf{L}(t)\cdot\boldsymbol{\xi}\right],1)^T$. 
We may then rewrite Eq.~\eqref{nevo} as the concise linear partial differential equation

\begin{equation}
	\mathbf{b}(\boldsymbol{\alpha})\cdot\nabla_{\boldsymbol{\alpha}}{P}(\boldsymbol{\alpha}) = 0
	.
\end{equation}
This type of partial differential equation can be solved using the method of characteristics. The characteristic solutions are parameterized by $s$ in the system of first order linear ordinary differential equations given by

\begin{subequations}
\begin{eqnarray}
	\frac{d\boldsymbol{\alpha}}{ds} = &&
	\mathbf{b}[\boldsymbol{\alpha}(s)]
\label{char1}
	,\\
	\frac{d{P}}{ds} = && 0
	.
\label{char2}
\end{eqnarray}
\end{subequations}
Eq.~\eqref{char1} is a vector equation with four components (three for $\boldsymbol{\xi}$, one for $t$). 
Consider the $t$ component,

\begin{equation}
	\frac{dt}{ds} = 
	1
	,
\end{equation}
which simply shows that $t$ and $s$ differ by a constant, which we will now choose to be zero, thereby taking $t(s)=s$. The $\boldsymbol{\xi}$ components of Eq.~\eqref{char1},

\begin{equation}
	\frac{d\boldsymbol{\xi}}{ds} = 
	\mathbf{L}(s)\cdot\boldsymbol{\xi}
	,
\end{equation}
are solved after taking $\boldsymbol{\xi}(0)=\boldsymbol{\xi}_0$ and $\mathbf{F}(0)=\mathbf{1}$ by

\begin{equation}
	\boldsymbol{\xi}(s) = 
	\mathbf{F}(s)\cdot\boldsymbol{\xi}_0
	.
\label{solxi}
\end{equation}
Eq.~\eqref{char2} shows that ${P}$ is constant when varying only $s$, 

\begin{equation}
	{P}\left[\boldsymbol{\xi}(s),t(s)\right] = 
	\mathrm{constant}
	,
\end{equation}
where using Eq.~\eqref{solxi} and $t=s$, and assuming we know ${P}$ at some previous time $\tau<t$, we then have the solution

\begin{equation}
	{P}\left[\mathbf{F}(t)\cdot\boldsymbol{\xi}_0,t\right] = 
	{P}\left[\mathbf{F}(\tau)\cdot\boldsymbol{\xi}_0,\tau\right]
	.
\end{equation}
We are free to choose $\boldsymbol{\xi}_0=\mathbf{F}^{-1}\cdot\boldsymbol{\xi}$ and retrieve

\begin{equation}
	{P}\left(\boldsymbol{\xi},t\right) = 
	{P}\left[{}_{(\tau)}\mathbf{F}^{-1}(t)\cdot\boldsymbol{\xi},\tau\right]
	,
\end{equation}
where we have used the deformation at $t$ relative to the deformation at a previous time previous time $\tau<t$, denoted as ${}_{(\tau)}\mathbf{F}(t)$, given by Paolucci \cite{paolucci2016continuum} as

\begin{equation}
	{}_{(\tau)}\mathbf{F}(t) = 
	\mathbf{F}(t)\cdot\mathbf{F}^{-1}(\tau)
	.
\end{equation}
Now, if we presume that the network is at equilibrium at time $\tau=0$, we have ${P}(\boldsymbol{\xi},0)={P}^\mathrm{eq}(\boldsymbol{\xi})$ and ${}_{(0)}\mathbf{F}(t)=\mathbf{F}(t)$, and our solution then becomes

\begin{equation}
	{P}(\boldsymbol{\xi},t) = 
	{P}^\mathrm{eq}\left[\mathbf{F}^{-1}(t)\cdot\boldsymbol{\xi}\right]
	,
\end{equation}
which is Eq.~\eqref{nsol} from the manuscript.

%%%%%%%%%%%%%%%%%%%%%%%%%%%%%%%%%%%%%%%%%%%%%%%%%%%%%%%%%%
\section{\label{appstress}Retrieving the stress}

Starting with Eqs.~\eqref{acon} and \eqref{stressineq}, we seek to retrieve a closed-form relation for the stress that does not include gradients of the network distribution of end-to-end vectors. 
We first relate the stress to the Helmholtz free energy using the hyperelastic formula (by way of neglecting dissipative stresses) and the solution for the distribution evolution in Eq.~\eqref{nevo}, where a spherical pressure term is included due to incompressibility. 
We then perform integration by parts, and after proving that the resulting boundary integral term is zero for relevant chain models, we retrieve the stress as an integral function of the network equilibrium distribution, the applied deformation, and the single chain mechanical response. 

%%%%%%%%%%%%%%%%%%%%%%%%%%%%%%%%%%%%%%%%%%%%%%%%%%%%%%%%%%
\subsection{\label{appsimpl}Simplifications and integration by parts}

We begin by taking the time derivative of $a$ given by Eq.~\eqref{acon} under constant temperature, so that it can later be used with Eq.~\eqref{stressineq},

\begin{eqnarray}
	\left(\frac{\partial a}{\partial t}\right)_T = &&
	n \iiint 
	\left(\frac{\partial {P}}{\partial t}\right)_T \psi^*
	 \,d^3\boldsymbol{\xi}
	-  p\left(\frac{\partial J}{\partial t}\right)_T
	,
\end{eqnarray}
where we know the evolution of $P$ from Eq.~\eqref{nevo}, we have chosen $n$ to be constant due to the incompressibility constraint (we are free to consider the $n$ derivatives to be nonzero and carry them through this derivation, but at the end they will be lumped into the pressure since they only produce spherical terms, thus leaving the same results).
For the last term, we use

\begin{equation}
	\left(\frac{\partial J}{\partial t}\right)_T = 
	J\mathbf{1}:\mathbf{L}
	.
\end{equation}
Substituting (with $J=1$) the above into Eq.~\eqref{stressineq} and neglecting dissipative stresses (taking the inequality to be an equality) then shows that the stress must be

\begin{equation}
	\boldsymbol{\sigma} = 
	-n\iiint \left(\frac{\partial {P}}{\partial\boldsymbol{\xi}}\right) \psi^* \boldsymbol{\xi} \,d^3\boldsymbol{\xi} - p\mathbf{1}
	.
\label{considerfirst}
\end{equation}
We now seek to rewrite the stress in a way that does not include gradients of ${P}$. 
We perform the integration by parts

\begin{eqnarray}
	\iiint \left(\frac{\partial {P}}{\partial\boldsymbol{\xi}}\right) \psi^* \boldsymbol{\xi} \,d^3\boldsymbol{\xi} = &&
	\iint {P} \psi^* \left(\frac{\boldsymbol{\xi}\boldsymbol{\xi}}{\xi}\right) \,d^2\boldsymbol{\xi} 
	\nonumber \\ 
	&&- \iiint {P}\left[\left(\frac{\partial \psi^*}{\partial\boldsymbol{\xi}}\right)\boldsymbol{\xi} + \psi^*\mathbf{1}\right] \,d^3\boldsymbol{\xi}
	,
	\nonumber\\ 
\end{eqnarray}
where the double integral is along the boundary $\xi\to\infty$ with unit normal vector $\hat{\boldsymbol{\xi}}=\boldsymbol{\xi}/\xi$, and $d^2\boldsymbol{\xi} $ is the surface element. 
We are free to lump the spherical term into $p\mathbf{1}$ without loss of generality.
Note that we have not taken into account that the molecular partition functions may depend on the volume -- this dependence would produce more spherical terms that would also now be lumped into $p$. 
The stress is now written as

\begin{eqnarray}
	\boldsymbol{\sigma} = &&
	n\iiint {P} \left(\frac{\partial \psi^*}{\partial\boldsymbol{\xi}}\right)  \boldsymbol{\xi} \,d^3\boldsymbol{\xi}
	%\nonumber \\ &&
	- n\iint {P} \psi^* \left(\frac{\boldsymbol{\xi}\boldsymbol{\xi}}{\xi}\right) \,d^2\boldsymbol{\xi} 
	 - p\mathbf{1}
	 .
	 \nonumber \\ 
\end{eqnarray}
If we then assume that $\psi^*$ is rotationally symmetric, we receive the non-polar stress 

\begin{eqnarray}
	\boldsymbol{\sigma} = &&
	n\iiint {P} \left(\frac{\partial\psi^*}{\partial\xi}\right)  \left(\frac{\boldsymbol{\xi}\boldsymbol{\xi}}{\xi}\right) \,d^3\boldsymbol{\xi}
	\nonumber \\ &&
	- n \iint {P} \psi^* \left(\frac{\boldsymbol{\xi}\boldsymbol{\xi}}{\xi}\right) \,d^2\boldsymbol{\xi} 
	 - p\mathbf{1}
	 .
\label{stresswithdp}
\end{eqnarray}
Taking the term from the integration along the boundary to be zero, taking $p=p^\mathrm{eq}+\Delta p$, and using the solution for $P$ in Eq.~\eqref{nsol}, we receive 

\begin{eqnarray}
	\boldsymbol{\sigma}(t) =&&
	n\iiint {P}^\mathrm{eq}\left[\mathbf{F}^{-1}(t)\cdot\boldsymbol{\xi}\right] \left(\frac{\partial\psi^*}{\partial\xi}\right) \left(\frac{\boldsymbol{\xi}\boldsymbol{\xi}}{\xi}\right) \,d^3\boldsymbol{\xi} 
	\nonumber \\ 
	&&- \left[p^\mathrm{eq} + \Delta p(t)\right]\mathbf{1}
	,
\end{eqnarray}
which is Eq.~\eqref{stressmixedrep} from the manuscript.
In the following section we discuss the boundary integral stress term in depth.

%%%%%%%%%%%%%%%%%%%%%%%%%%%%%%%%%%%%%%%%%%%%%%%%%%%%%%%%%%
\subsection{\label{appIBP}The boundary integral stress term}

We now consider the stress from the integration along the boundary in Eq.~\eqref{stresswithdp} in order to show that it is zero for arbitrary deformations as long as $\psi^*$ satisfies certain growth criteria. 
Using Eq.~\eqref{nsol} and taking $d\boldsymbol{\mathcal{S}} = (\boldsymbol{\xi}/\xi)\,d^2\boldsymbol{\xi}$, the boundary integral stress term is

\begin{equation}
	\hat{\boldsymbol{\sigma}} = 
	n \iint {P}^\mathrm{eq}\left(\mathbf{F}^{-1}\cdot\boldsymbol{\xi}\right) \psi^*(\boldsymbol{\xi}) \boldsymbol{\xi}\,d\boldsymbol{\mathcal{S}}
	.
\end{equation}
We recall that $\psi^*$ had been assumed to be rotationally symmetric, and the boundary is along $\xi\to\infty$, so we may take $\psi^*$ out of the integrand and write

\begin{equation}
	\hat{\boldsymbol{\sigma}} = 
	n\lim_{\xi\to\infty}\left[\psi^*(\xi)\right] \iint {P}^\mathrm{eq}\left(\mathbf{F}^{-1}\cdot\boldsymbol{\xi}\right) \boldsymbol{\xi}\,d\boldsymbol{\mathcal{S}}
	.
\end{equation}
We take the change of variables $\boldsymbol{\xi}\mapsto\mathbf{F}\cdot\boldsymbol{\xi}$, where as pointed out by Paolucci \cite{paolucci2016continuum}, the surface element transforms as $d\boldsymbol{\mathcal{S}}\mapsto J\mathbf{F}^{-T}\cdot d\boldsymbol{\mathcal{S}}$.
We also use $J=1$ and then receive

\begin{equation}
	\hat{\boldsymbol{\sigma}} = 
	n \lim_{\xi\to\infty} \left[\psi^*(\xi)\right] \iint {P}^\mathrm{eq} \left(\boldsymbol{\xi}\right) \left(\mathbf{F}\cdot\boldsymbol{\xi}\right)\mathbf{F}^{-T}\cdot d\boldsymbol{\mathcal{S}}
	.
\end{equation}
Since $\psi^*$ is rotationally symmetric, by the distribution-behavior correspondence in Eq.~\eqref{equiv1} $P^\mathrm{eq}$ is rotationally symmetric as well.
We may then remove $P^\mathrm{eq}$ from the integrand along with the deformation terms for

\begin{equation}
	\hat{\boldsymbol{\sigma}} = 
	n\lim_{\xi\to\infty}\Big[\psi^*(\xi){P}^\mathrm{eq}(\xi)\Big] \left\{\mathbf{F}\mathbf{F}^{-T}:\iint \boldsymbol{\xi}\,d\boldsymbol{\mathcal{S}}\right\}
	,
\end{equation}
and now we use $d\boldsymbol{\mathcal{S}}=\xi\boldsymbol{\xi}\,d\Omega$, where $d\Omega$ is the differential solid angle, and remove factors of $\xi$ from the integrand to finally write $\hat{\boldsymbol{\sigma}}$ as

\begin{equation}
	\hat{\boldsymbol{\sigma}} = 
	\left(\frac{4\pi n}{3}\,\mathbf{1} \right)\lim_{\xi\to\infty} \Big[\xi^3\psi^*(\xi){P}^\mathrm{eq}(\xi)\Big]
\label{limiteqn}
	,
\end{equation}
where the term in the parentheses was retrieved from

\begin{equation}
	\mathbf{F}\mathbf{F}^{-T}:\iint \hat{\boldsymbol{\xi}}\hat{\boldsymbol{\xi}}\,d\Omega = 
	\mathbf{F}\cdot\left(\frac{4\pi}{3}\,\mathbf{1}\right)\cdot\mathbf{F}^{-1} = 
	\frac{4\pi}{3}\,\mathbf{1}
	.
\end{equation}
So far we have shown that the stress contributed by $\hat{\boldsymbol{\sigma}}$ is spherical, and could therefore be lumped into the pressure term $-p\mathbf{1}$, as long as it is finite.
We are then tasked with proving that the limit in Eq.~\eqref{limiteqn} is finite, but we will instead prove it is zero.
We will accomplish this by requiring that $\psi^*$ grows sufficiently fast as $\xi\to\infty$, and using l'H\^{o}pital's rule and the squeeze theorem. 
To start, we require that the growth of $\psi^*$ as $\xi\to\infty$ is greater than that of a logarithm function, i.e.

\begin{equation}
	\lim_{\xi\to\infty} \left[\frac{\psi^*(\xi)}{c\ln(\xi)}\right] = \infty
	,\qquad \forall c > 0
\label{requirementpsi}
	.
\end{equation}
This is required to guarantee that the denominator of Eq.~\eqref{equiv1} is finite, but we will use it here as well.
Take Eq.~\eqref{equiv1} with the denominator now understood to be a finite constant, thus eliminating it, as we use the limit in Eq.~\eqref{limiteqn} to define the functional of $\psi^*(\xi)$,

\begin{equation}
	W(\psi^*) = 
	\lim_{\xi\to\infty} \left[\frac{\xi^3\psi^*(\xi)}{e^{\beta\psi^*(\xi)}}\right]
\label{functionallimit}
	,
\end{equation}
where we can assume that $\beta>0$. 
See from Eq.~\eqref{requirementpsi} that we have already required that $\psi^*$ is positive as $\xi\to\infty$, so we have in turn required that $W(\psi^*) \geq 0$. 
Also see from Eq.~\eqref{functionallimit} that $W(\psi^*) \leq W(c\ln\xi)$ for all $\psi^*$ that satisfy the requirement from Eq.~\eqref{requirementpsi}, so together we have
\begin{equation}
	W(c\ln\xi) \geq W(\psi^*) \geq 0
	,\qquad \forall c > 0
\label{squeezeeqn}
	.
\end{equation}
Next, let us consider the special (albeit prohibited for admissible $\psi^*$) case of $\psi^*=c\ln\xi$, where we repeatedly use l'H\^{o}pital's rule to show

\begin{eqnarray}
	W(c\ln\xi) = &&
	\lim_{\xi\to\infty} \left[\frac{c\xi^3\ln\xi}{\xi^{c\beta}}\right] 
	,\\ 
	= &&
	\lim_{\xi\to\infty} \left[\frac{6 \xi^{3 - c\beta}}{\beta(c\beta - 1)(c\beta - 2)(c\beta - 3)}\right]
	,~~~~
\end{eqnarray}
in order to see that $W(c\ln\xi) = 0$ for all $c\beta>3$. 
Since Eq.~\eqref{squeezeeqn} holds for all $c\beta>3$, by the squeeze theorem we retrieve $W(\psi^*)=0$ for any $\psi^*$ that satisfies Eq.~\eqref{requirementpsi}.
Thus, the stress from the integration along the boundary must be zero, $\hat{\boldsymbol{\sigma}} = \mathbf{0}$, for any rotationally symmetric $\psi^*$ that grows faster than logarithmically as $\xi\to\infty$. 
This is true for the EFJC model considered here. 

%%%%%%%%%%%%%%%%%%%%%%%%%%%%%%%%%%%%%%%%%%%%%%%%%%%%%%%%%%
\section{\label{neohook}Reduction to Neo-Hookean model}

The Neo-Hookean model is retrieved when the ideal chain free energy in Eq.~\eqref{psigaussK} and the corresponding equilibrium distribution in Eq.~\eqref{PgaussK} are utilized. 
Substitution of these into the stress from Eq.~\eqref{stressmixedrep} yields

\begin{eqnarray}
	\boldsymbol{\sigma} = &&
	\frac{nkT}{(2\pi)^{3/2}} \iiint \exp\left(-\frac{1}{2}\left\|\mathbf{F}^{-1}\cdot\mathbf{v}\right\|_2^2\right) \mathbf{v}\mathbf{v}\,d^3\mathbf{v}
	\nonumber \\  &&
	- \left[p^\mathrm{eq} + \Delta p\right]\mathbf{1}
	,
\end{eqnarray}
where we have made convenient use of the new non-dimensional variable

\begin{equation}
\mathbf{v} 
 = \boldsymbol{\lambda}\sqrt{3c_\kappa N_b}
 = \boldsymbol{\xi}\sqrt{\frac{3c_\kappa}{N_b\ell_b^2}}
,
\end{equation}
and where $\|\mathbf{a}\|_2^2=\mathbf{a}\cdot\mathbf{a}$. 
We may take the change of variables $\mathbf{v}\mapsto\mathbf{F}\cdot\mathbf{v}$, where $d^3\mathbf{v}$ is unaltered due to incompressibility, to instead write the stress as

\begin{eqnarray}
	\boldsymbol{\sigma} = &&
	\frac{nkT}{(2\pi)^{3/2}}\left(\mathbf{F}\mathbf{F}^T:\iiint e^{-v^2/2}\, \mathbf{v}\mathbf{v}\,d^3\mathbf{v}\right)
	\nonumber \\  &&
	- \left[p^\mathrm{eq} + \Delta p\right]\mathbf{1}
	.
\end{eqnarray}
This integral can be directly computed,

\begin{equation}
\iiint e^{-v^2/2}\, \mathbf{v}\mathbf{v}\,d^3\mathbf{v} = (2\pi)^{3/2}\,\mathbf{1}
,
\end{equation}
and we compute the same type of integral in order to retrieve $p^\mathrm{eq}=nkT$.
Our end result is now seen to be

\begin{equation}
	\boldsymbol{\sigma}(t) =
	nkT\left[\mathbf{B}(t) - \mathbf{1}\right] - \Delta p(t)\mathbf{1}
	,
\end{equation}
which is exactly the incompressible Neo-Hookean model with shear modulus $nkT$.

%%%%%%%%%%%%%%%%%%%%%%%%%%%%%%%%%%%%%%%%%%%%%%%%%%%%%%%%%%
\section{\label{appsmall}Infinitesimal deformation}

Here we seek to find the reduced form of the stress from Eq.~\eqref{stressmixedrep} when an infinitesimal deformation $\mathbf{F}=\mathbf{1}+\mathbf{E}$ is applied, where $\mathbf{E}^2\approx 0$. 
Incompressibility is now enforced via $\mathrm{tr}(\mathbf{E})=0$, and the inverse is $\mathbf{F}^{-1}=\mathbf{1}-\mathbf{E}$. 
We take $P^\mathrm{eq}(\mathbf{F}^{-1}\cdot\boldsymbol{\xi})$ appearing in Eq.~\eqref{stressmixedrep}, which we can write in term of $\psi^*(\mathbf{F}^{-1}\cdot\boldsymbol{\xi})$ using Eq.~\eqref{equiv1}, and is expanded as

\begin{equation}
	\psi^*(\mathbf{F}^{-1}\cdot\boldsymbol{\xi}) = 
	\psi^*(\boldsymbol{\xi}) - \mathbf{E}:\left(\frac{\partial \psi^*}{\partial\boldsymbol{\xi}}\,\boldsymbol{\xi}\right) + O(\mathbf{E}^2)
	.
\end{equation}
We substitute this into Eq.~\eqref{equiv1} to obtain

\begin{equation}
	P^\mathrm{eq}(\mathbf{F}^{-1}\cdot\boldsymbol{\xi}) = 
	P^\mathrm{eq}(\boldsymbol{\xi}) e^{\mathbf{E}:\left(\frac{\partial\beta\psi^*}{\partial\boldsymbol{\xi}}\,\boldsymbol{\xi}\right) + O(\mathbf{E}^2)}
	.
\end{equation}
The equilibrium pressure $p^\mathrm{eq}$ is the form of the stress when $\mathbf{F}=\mathbf{1}$, so we will receive a term in the integrand of the following form, which is simplified for small $\mathbf{E}$ as

\begin{eqnarray}
	&& P^\mathrm{eq}(\boldsymbol{\xi})\left[e^{\mathbf{E}:\left(\frac{\partial\beta\psi^*}{\partial\boldsymbol{\xi}}\,\boldsymbol{\xi}\right) + O(\mathbf{E}^2)} - 1\right]
	\nonumber \\ &&~~~~=
	P^\mathrm{eq}(\boldsymbol{\xi})\left[\mathbf{E}:\left(\frac{\partial\beta\psi^*}{\partial\boldsymbol{\xi}}\,\boldsymbol{\xi}\right) + O(\mathbf{E}^2)\right]
	,\\ &&~~~~=
	\mathbf{E}:\left(-\frac{\partial P^\mathrm{eq}}{\partial\boldsymbol{\xi}}\,\boldsymbol{\xi}\right) + O(\mathbf{E}^2)
	.
\end{eqnarray}
We now recall that both $P^\mathrm{eq}$ and $\psi^*$ are spherically symmetric, and substitute into Eq.~\eqref{stressmixedrep} for

\begin{eqnarray}
	\boldsymbol{\sigma} = &&
	n\int_0^\infty \left(-\frac{\partial P^\mathrm{eq}}{\partial\xi}\right)\left(\frac{\partial\psi^*}{\partial\xi}\right)\xi^4\,d\xi
	\nonumber \\ && \times
	\mathbf{E}:\iint \hat{\boldsymbol{\xi}}\hat{\boldsymbol{\xi}}\hat{\boldsymbol{\xi}}\hat{\boldsymbol{\xi}} \,d\Omega
	- \Delta p\mathbf{1} + O(\mathbf{E}^2)
	.
\end{eqnarray}
Now, after completing the integral

\begin{equation}
	\mathbf{E}:\iint \hat{\boldsymbol{\xi}}\hat{\boldsymbol{\xi}}\hat{\boldsymbol{\xi}}\hat{\boldsymbol{\xi}} \,d\Omega = 
	\frac{4\pi}{15}\left[2\mathbf{E} + \mathrm{tr}(\mathbf{E})\mathbf{1}\right]
	,
\end{equation}
where $\mathrm{tr}(\mathbf{E})=0$ here, and then defining the shear modulus as

\begin{equation}
	\mu = 
	\frac{8\pi}{15}\,nkT\iiint \left(-\frac{\partial{P}^\mathrm{eq}}{\partial\xi}\right)\left(\frac{\partial\beta\psi^*}{\partial\xi}\right)\xi^4\,d\xi
	,
\end{equation}
we can finally write the stress as

\begin{equation}
	\boldsymbol{\sigma}(t) =
	\mu\mathbf{E}(t) - \Delta p(t)\mathbf{1} + O(\mathbf{E}^2)
	,
\end{equation}
which are Eqs.~\eqref{stress_small} and \eqref{modulus_small} from the manuscript. 

%%%%%%%%%%%%%%%%%%%%%%%%%%%%%%%%%%%%%%%%%%%%%%%%%%%%%%%%%%
%%%%%%%%%%%%%%%%%%%%%%%%%%%%%%%%%%%%%%%%%%%%%%%%%%%%%%%%%%%%%%%%%%%%%%%%%%%%%%%%%%%%

% The \nocite command causes all entries in a bibliography to be printed out
% whether or not they are actually referenced in the text. This is appropriate
% for the sample file to show the different styles of references, but authors
% most likely will not want to use it.
% \nocite{*}

\bibliography{main}% Produces the bibliography via BibTeX.

\end{document}